\documentclass[preprint,12pt]{elsarticle}
\usepackage[a4paper, left=2cm, right=2cm, top=2.5cm, bottom=2.5cm]{geometry}

\usepackage{graphicx}
\usepackage{dcolumn}
\usepackage{bm}
\usepackage{amssymb}
\usepackage{amsmath}
\usepackage{color}
\usepackage{physics}
\biboptions{sort&compress}
\usepackage{soul}
\usepackage{hyperref}

\hypersetup{
    colorlinks=true, linkcolor=blue, citecolor=blue,
    filecolor=blue, urlcolor=blue, breaklinks=true
}

\newcommand{\orcid}[1]{\href{https://orcid.org/#1}
  {\includegraphics[width=7pt]{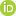}}}

\begin{document}

\title{Climate and dengue synchronization in southern Brazil: a municipal analysis with cross-state validation
}

\author{
Enrique C. Gabrick$^{1,6}$, Antonio M. Batista$^{2,3}$, Iber\^e L. Caldas$^{1}$, J\"urgen Kurths$^{4,5}$, Ma\'ira Aguiar$^{6,7}$}

\address{$^{1}$Institute of Physics, University of S\~ao Paulo,  S\~ao Paulo, Brazil.\\
$^{2}$Graduate Program in Science, State University of Ponta Grossa, Ponta Grossa, Brazil. \\
$^{3}$Department of Mathematics and Statistics, State University of Ponta Grossa, Ponta Grossa, Brazil. \\
$^{4}$Potsdam Institute for Climate Impact Research, Potsdam, Germany. \\
$^{5}$Department of Physics, Humboldt University Berlin, Berlin, Germany. \\
$^{6}$BCAM - Basque Center for Applied Mathematics, Bilbao, Spain. \\
$^{7}$Ikerbasque, Basque Foundation for Science, Bilbao, Spain.
}


\cortext[cor]{ecgabrick@gmail.com}

\begin{abstract}
Dengue transmission is rapidly expanding beyond its historical tropical range, raising concerns about how climate change may alter the collective dynamics of epidemics. While most studies focus on transmission risk, much less is known about how climate affects the synchronization of outbreaks. In this work, we investigate dengue synchronization using epidemiological and climate data from 74 municipalities in the state of Paraná (southern Brazil) between 2010 and 2024. We quantify outbreak coherence using the Event Synchronization (ES) method. Our results reveal a transition from a low-transmission regime to a high-transmission regime accompanied by a marked increase in synchronization across cities. We also show that climate anomalies increase the number of permissive days for dengue transmission. Our results suggest that such days are significantly associated with outbreak synchronization. We identify a two-stage climate mechanism: conducive climatic conditions first reduce the probability of asynchronous states and coincide with the emergence of synchronized outbreaks, and subsequently sustain higher synchronization levels. Extending the analysis through comparative analyses in Ceará and Minas Gerais, we uncover that climate consistently amplifies synchronization, although its role in the onset of synchronization depends on regional climatic regimes. These findings highlight climate-driven synchronization as an emerging feature shaping dengue dynamics.
\end{abstract}


\maketitle
\sloppy
\section{Introduction}
Transmitted by {\it Aedes aegypti} and {\it Aedes albopictus} mosquitoes, dengue now threatens about half of the world's population, with an estimated 100-400 million infections each year \cite{Who2025,Bhatt2013}. Dengue is caused by four distinct viral serotypes (DENV-1 to DENV-4) and is spread to humans through the bites of infected mosquitoes \cite{Simmons2012}. Most cases are asymptomatic or mild, and infected individuals typically recover within 1-2 weeks. However, the immune response to dengue is complex \cite{Kularatne2022}. Infection with one serotype provides long-term immunity to that serotype and temporary cross-protection against the others \cite{Dash2024}. Once this cross-protection wanes, subsequent infections with a different serotype increase the risk of severe dengue, which can be fatal \cite{Guzman2015}. However, recent findings suggest that complete lifelong protection against the same serotype may not always hold, highlighting the need to reconsider dengue immunity and reinfection dynamics \cite{Afrina21,Anam2024}.

Mathematical modelling has been central to dengue research, providing insight into how multi-serotype transmission, temporary cross-immunity, antibody-dependent enhancement, vaccination, vector dynamics, and control interventions shape epidemic behaviour. Previous models have shown that dengue can generate complex nonlinear dynamics, including oscillations, bifurcations, chaotic behaviour, and sensitivity to data quality and parameter estimation \cite{Aguiar2010,Aguiar2012,Aguiar2013,Aguiar2014,Steindorf2022,Aguiar2022}. Other studies have examined dengue vaccination, serotype-specific cross-protection, and the interpretation of vaccine trials and public-health interventions \cite{Aguiar2016,Aguiar2018,Aguiar2020}. More recent modelling frameworks have extended these approaches to explicit vector dynamics and behavioural or control mechanisms, highlighting the importance of integrating biological, epidemiological, and environmental processes in dengue models \cite{Srivastav2026,Doutor2025}. While these studies have substantially advanced the understanding of dengue transmission mechanisms, less attention has been given to how climate variability may shape the synchronization of outbreaks across municipalities.

Because dengue transmission depends on the interaction between host immunity, mosquito ecology, and environmental suitability, understanding the climatic sensitivity of the vector is essential \cite{Aguiar2011,Rocha2014}. Mosquitoes are directly affected by climate and environmental conditions \cite{Gabrick2025,Merkenschlager2025,Gibb2023}. Temperature substantially modulates dengue transmission by regulating mosquito development, survival, biting rates, and the virus's extrinsic incubation period \cite{Abdullah2022}. These features respond strongly to temperature and are amplified within the range $18\,^\circ$C and $34\,^\circ$C for \textit{Aedes aegypti} and \textit{Aedes albopictus}. However, the optimal efficiency within a narrower range \cite{Mordecai2017}.  On the other hand, precipitation exerts a more complex and non-linear influence on vector abundance \cite{Lowe2018,Liu2025}. Moderate rainfall regimes generally increase vector abundance by increasing the availability of breeding sites, whereas extreme rainfall can lead to flushing effects, as observed in Singapore \cite{Benedum2018}. Paradoxically, drought periods can also increase vector abundance, because drought can lead households to store water, which can create additional breeding sites \cite{Islam2019}. Due to the environmental conditions required for mosquito development and survival, dengue was historically confined to tropical and subtropical regions \cite{Bhatt2013}. However, recent studies indicate that climate change and other anthropogenic factors have contributed to the geographic expansion of dengue transmission risk beyond these regions \cite{ECDC2025,ECDC2025b,Souza2024b,Nakase2024,Estallo2023,Gonzalez2021}. This expansion is increasingly relevant in temperate settings, where invasive mosquitoes are becoming established \cite{Stollenwerk2025,Guerrero2025,Pisaneschi2026} and climate-based forecasting approaches have highlighted the role of temperature, precipitation, and humidity in shaping mosquito abundance and local transmission potential \cite{Steindorf2025}.

Climate change is one of the main factors linked to the changing geographical range of dengue and other infectious diseases \cite{Baker2022,Metcalf2017}. By altering vector distributions, behaviour, and density, favourable climatic conditions can contribute to outbreaks in previously unaffected regions, including dengue and chikungunya outbreaks in Europe \cite{Charnley2025,Tegar2026,Pereira2025}. However, warming can also exceed the upper thermal limits of mosquitoes and reduce transmission in some regions, such as North Africa \cite{Ryan2019}. Human mobility further accelerates the spread of vectors and pathogens across regions \cite{Kraemer2019}. Although the expansion of dengue transmission risk under climate change has been widely studied \cite{Barcellos2024,Lowe2025,Childs2024,Cox2025}, much less is known about how climate variability affects the collective spatio-temporal dynamics of dengue outbreaks across regions.


Understanding collective epidemic patterns requires tools capable of quantifying coordinated dynamics between locations. One such approach is the use of synchronization metrics, which are widely employed in several fields where oscillating systems adjust their rhythms \cite{KurthsBook}. In epidemiological systems, synchronization provides a natural framework to quantify the coherence of outbreaks across spatially separated regions \cite{Viboud2006}. Analysing dengue surveillance data from 14 countries across the Americas, Johansson et al. reported strong synchronization of dengue dynamics at both seasonal and multiannual timescales. Large epidemics at multiannual scales were shared across the region, with an average temporal lag of only six months even between locations separated by distances of up to 10,000 km. The authors also showed that seasonal synchrony is associated with climatic variables, including temperature, rainfall, and El Niño Southern Oscillation \cite{Johansson2025}. Similar region-wide synchrony and travelling-wave behaviour have also been documented across Southeast Asia, where elevated temperatures were linked to synchronous dengue epidemics \cite{Panhuis2015}. Results from Thailand likewise indicate that dengue synchrony is related to temperature patterns: periods of strong synchronization coincide with coherent temperature fluctuations, whereas multiannual oscillations associated with population immunity can generate asynchronous dynamics \cite{Carreras2022,Schwartz2005}. Under such conditions, temperature synchrony can act as an external driver capable of re-synchronizing dengue epidemics across regions \cite{Carreras2022}. In Brazil, dengue epidemics exhibit a clear spatio-temporal structure, with travelling waves typically starting in the western states and propagating eastward. Consistent with this spatial propagation, epidemic synchrony decreases with increasing geographical distance between regions. These spatial patterns are shaped by a combination of human mobility, supported by gravity-style models, and climate factors \cite{Churakov2019}. Although dengue synchrony has been linked to climate variability in several regions worldwide, much less is known about synchronization among cities in Brazil, one of the countries most heavily affected by dengue.

In this work, we investigate how dengue synchronization is modulated by climatic factors, particularly temperature and precipitation. Although synchronization in epidemic systems has been extensively studied \cite{Rohani1999,Earn1998,Rodriguez2023,Cummings2004}, most existing approaches rely on wavelet-based methods \cite{Johansson2025,Carreras2022} or pairwise correlation analyses \cite{Churakov2019}. Here, we quantify outbreak synchronization using an adaptation of the nonlinear event synchronization (ES) method \cite{Quiroga2002,Quiroga2002b}, a framework for measuring the temporal alignment of discrete events, such as rainfall. ES has been widely employed to quantify nonlinear interdependence in complex systems, with applications in neuroscience \cite{Kreuz2007,Pereda2005} and climate science \cite{Tang2025,Boers2019,Zhen2022}. To the best of our knowledge, this method has not yet been systematically applied to characterize synchronization in epidemic outbreaks. Its application therefore represents a novel methodological contribution to the study of infectious disease dynamics. We apply ES to dengue surveillance data from municipalities in the state of Paran\'a (southern Brazil) between 2010 and 2024, a region where endemic and non-endemic cities coexist. Our analysis reveals a clear epidemiological transition characterized by the geographical expansion of dengue transmission and increasing temporal coherence of outbreaks across cities. We further show that anomalous climatic conditions modify the number of climate-permissive days for transmission and are statistically associated with changes in outbreak synchronization. Using linear and hurdle regression models, we assess how these climatic variables relate to synchronization. The linear model reveals a positive and statistically significant association between the number of conducive days and increasing synchronization. The hurdle model suggests a two-stage mechanism linking climate variability to epidemic synchronization. First, favourable climatic conditions are associated with a reduced probability of asynchronous transmission and coincide with the emergence of synchronized outbreaks. Once synchronization is established, dynamically adequate climatic conditions are associated with higher and more persistent coherent epidemic activity. To assess the generality of this pattern, we extend the analysis to two additional endemic states in Brazil, Ceará and Minas Gerais, revealing that while the amplification pattern is robust, the role of climate in the onset of synchronization depends on the regional climatic regime.

This study aims to quantify synchronization across municipalities, test whether synchronization changed after 2020, link synchronization to climate-permissive thermal and rainfall conditions, and assess whether similar patterns appear in other Brazilian states. This paper is organized as follows. In Section \ref{methodology}, we describe the epidemiological and climatic datasets together with the event synchronization methodology. Section \ref{phase_transition} presents evidence of a phase transition associated with the geographical expansion of dengue transmission. In Section \ref{synchronization}, we analyse the spatio-temporal synchronization patterns of dengue outbreaks. Section \ref{anomalia_sec} investigates how climatic anomalies modify the number of climate-permissive transmission days and how these variables relate to synchronization dynamics. In Section \ref{validation}, we extend the analysis to two additional Brazilian states to evaluate the generality of the proposed pattern. In Section \ref{sensitivity_analysis}, we perform a sensitivity analysis of the synchronization window and outbreak threshold.
Finally, Section \ref{conclusions} summarizes the main findings and discusses their epidemiological implications.

\section{Methodology}\label{methodology}
\subsection{Dengue data}
In this work, we use notified dengue case data from the InfoDengue platform \cite{Infodengue}, an early-warning surveillance system that compiles arboviral (dengue, chikungunya, and Zika) case notifications for all Brazilian municipalities. The platform integrates epidemiological, climate, and demographic information from multiple official sources and applies statistical and nowcasting procedures to correct reporting delays, generating a harmonized dataset with weekly resolution.

We analysed weekly dengue notifications from the first epidemiological week of 2010 to the last epidemiological week of 2024 (2010-01-03 to 2024-12-22), comprising 782 consecutive weeks. The dataset provides notified cases but does not include detailed serological confirmation data, as only a fraction of reported cases are laboratory-confirmed in Brazil. To describe the co-circulation of dengue virus serotypes over time, we obtained the data from the official state epidemiological reports \cite{Informeparana}.

Brazil is endemic for dengue and accounts for a substantial proportion of global cases annually. While the highest historical burden has been concentrated in the North, Northeast, and Central-West regions, southern states have traditionally reported lower incidence. However, they exhibit marked intra-state heterogeneity, with some municipalities classified as endemic and others presenting sporadic transmission. Paran\'a represents a paradigmatic example of this transitional epidemiological setting.

Paran\'a is characterized by pronounced climatic heterogeneity. The southern region has a temperate subtropical climate with cooler winters and mean summer temperatures below 22~$^\circ$C, whereas the northern and northwestern regions present warmer subtropical conditions, with mean summer temperatures exceeding 22~$^\circ$C. The coastal area exhibits a humid tropical climate. This climatic variability contributes to heterogeneous ecological suitability for dengue transmission across the state.

Climatic heterogeneity, combined with socioeconomic differences and population distribution across 399 municipalities, results in substantial spatial variability in dengue incidence. Municipal population sizes range from approximately 1,300 to 1.9 million inhabitants (IBGE census data from 2022 \cite{IBGE2022}). To reduce stochastic effects associated with very small populations and sporadic reporting, we applied two inclusion criteria: (i) municipalities with more than 20,000 inhabitants and (ii) municipalities reporting more than 1,000 cumulative dengue cases over the study period (2010--2024). After applying these filters, 74 municipalities were retained for our analysis, ensuring representation from all geographic regions of Paran\'a.

Although demographic data are available through InfoDengue, we used official population estimates from the 2022 Brazilian national census (IBGE) to ensure consistency and accuracy \cite{IBGE2022}.

Additionally, we ordered the municipalities by latitude and assigned a numerical index (IDX$ = 1,2,\ldots,74$).

\subsection{Climate data}
Although InfoDengue provides weekly climate indicators (e.g., mean temperature and humidity), these variables contain missing values for several municipalities in our dataset. To ensure spatial completeness and to construct a consistent historical baseline, we used the ERA5-Land reanalysis dataset \cite{ERA5land}. The dataset provides hourly surface climate variables with a horizontal resolution of $0.1^\circ \times 0.1^\circ$ (native resolution of 9 km), which is adequate for a municipality-level analysis.

For each municipality, we extracted the daily near-surface air temperature and total precipitation from 1980-01-01 to 2024-12-28. Municipal boundaries were obtained from the official 2024 Brazilian municipal grid provided by IBGE \cite{malhamunicipal}. Gridded climate data were spatially aggregated within each municipal boundary. Daily values were subsequently aggregated to weekly or monthly resolution when required for the analysis.

To quantify climate anomalies, we divided the dataset into a historical baseline period (1980-01-01 to 2010-01-02) and a study period (2010-01-03 to 2024-12-28). For temperature ($T$, in $^\circ$C) and precipitation ($P$, in mm), we computed the historical climatological mean for each month and municipality, denoted by $\overline{x}^{\rm{hist}}_{m,i}$, where $x \in \{T, P\}$, $m$ denotes the month, and $i$ denotes the municipality index. Then, we calculated the monthly anomalies, defined as deviations from the historical climatological reference, through
\begin{equation}
x^{an.}_{m,i} = x_{m,i} - \overline{x}^{\rm hist.}_{m,i}. \label{anomalia}
\end{equation}
where $x_{m,i}$ represents the observed monthly mean temperature or accumulated precipitation for municipality $i$ in month $m$ during the study period. Positive values indicate warmer or wetter conditions relative to the reference interval (1980-2010).

\subsection{Synchronization metric}
To quantify the temporal alignment of dengue outbreaks across municipalities, we employed the event synchronization (ES) method \cite{Quiroga2002,Quiroga2002b}, which measures the temporal coincidence of discrete events within a predefined tolerance window \cite{Agarwal2017}. ES is a robust metric that can be employed for time series characterized by intermittency, non-Gaussian distributions, and outbreak-like dynamics \cite{Boers2019}. This makes the method suitable for epidemic time series.

For each municipality $i$, outbreak events were identified from the weekly dengue case time series $c_i(t)$. First, we computed the mean weekly incidence over the entire study period, denoted by $\overline{c}_i$. An outbreak episode was defined as a period during which $c_i(t) \geq \overline{c}_i + 1$, beginning when $c_i(t)$ crossed above this threshold. The episode was considered terminated when the time series remained below this threshold for at least two consecutive weeks, preventing short-term fluctuations from being classified as separate outbreaks.

Within each outbreak interval $[t_{\rm start},\, t_{\rm end}]$, we defined a single event corresponding to the week of maximum incidence, according to
\begin{equation}
t^i_k = {\rm arg}\,{\rm max}_{t \in [t_{\rm start}, \, t_{\rm end}]} c_i(t),
\end{equation}
where $t^i_k$ represents the timing of the $k$-th outbreak peak in municipality $i$. Each outbreak interval, therefore, contributed one discrete event. The full set of outbreak peaks for municipality $i$ is given by $\{t^i_k\}_{k=1}^{N_i}$, where $N_i$ denotes the total number of outbreaks identified during the study period.

For two municipalities $x$ and $y$, with event series $\{t^x_k\}_{k=1}^{N_x}$ and $\{t^y_j\}_{j=1}^{N_y}$, respectively, two outbreak peaks were considered temporally synchronized when their occurrence times differed by at most a tolerance window $\tau$, that is,
\begin{equation}
|t^x_k - t^y_j| \leq \tau.
\end{equation}
Throughout this study, we adopted $\tau = 2$ weeks, which accounts for reporting delays and short temporal shifts in outbreak peaks while preserving epidemiologically meaningful coincidence.

To characterize synchronization at the state level, we constructed a weekly synchronization index ${\rm Syn}(t)$. For each municipality $i$ and week $t$, we defined
\begin{eqnarray}
E_i(t) =
\begin{cases}
1, & \text{if an event occurs in } [\,t-\tau,\, t+\tau\,], \\[6pt]
0, & \text{otherwise}.
\end{cases}
\end{eqnarray}
Then, ${\rm Syn}(t)$ is
\begin{eqnarray}
{\rm Syn}(t) =
\begin{cases}
\frac{1}{N} \sum_{i=1}^{N} E_i(t), & \text{if } \sum_{i=1}^{N} E_i(t) > 1, \\[6pt]
0, & \text{otherwise}, \label{synchronization_eq}
\end{cases}
\end{eqnarray}
where $N$ is the total number of municipalities and ${\rm Syn}(t) \in [0,1]$.

Therefore, ${\rm Syn}(t)$ represents the fraction of municipalities experiencing outbreak peaks within a $\pm 2$-week window centred at week $t$. Low values of ${\rm Syn}(t)$ indicate temporally localized transmission restricted to a small number of municipalities, whereas high values reflect concurrent outbreaks across a large portion of the state. In this framework, synchronization captures the degree of regional epidemic coherence rather than simply the magnitude of incidence.

\begin{figure}[!ht]
	\centering
	\includegraphics[scale=0.4]{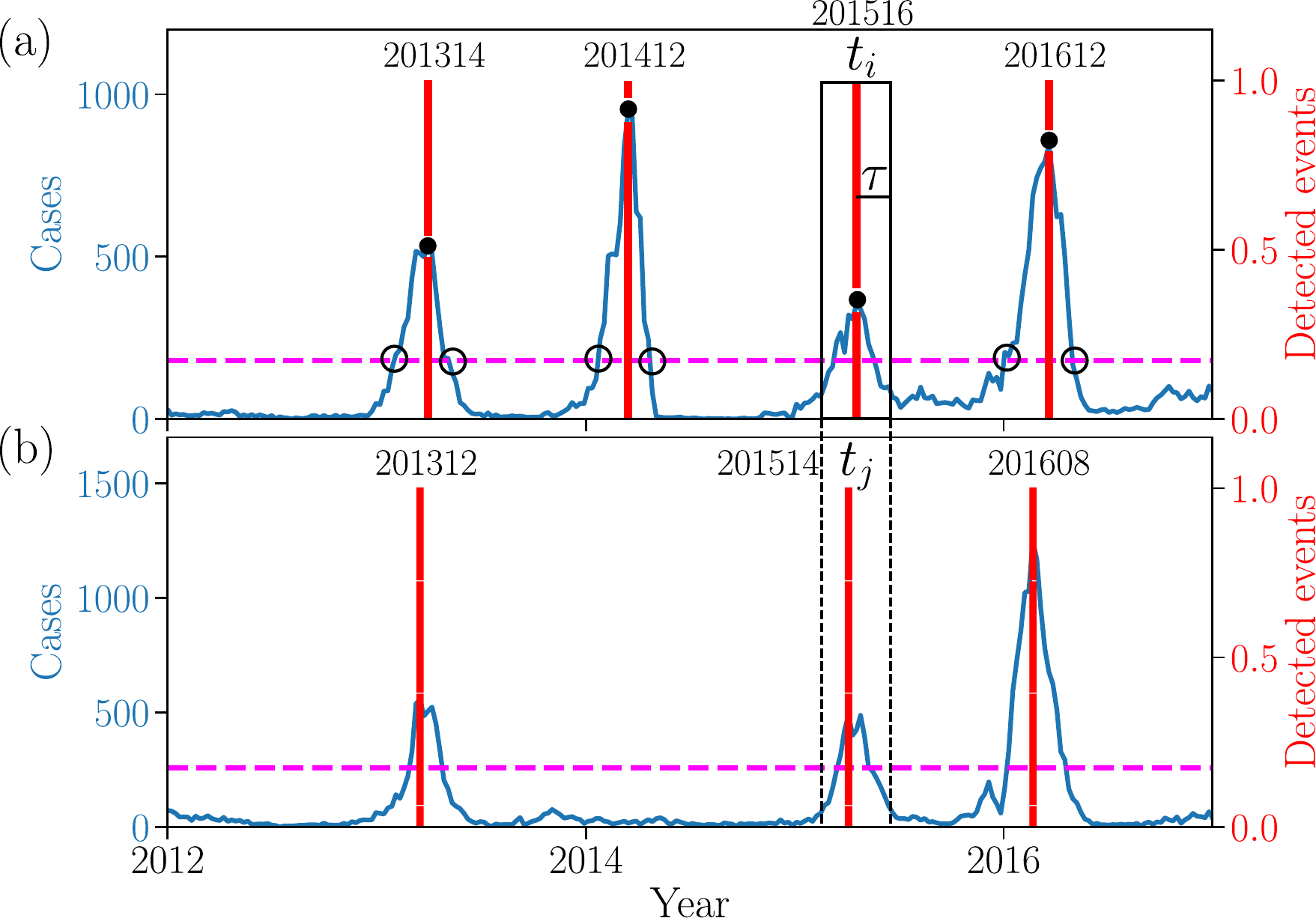}
	\caption{Illustration of outbreak detection and event synchronization for two municipalities: Maring\'a (a) and Foz do Iguaçu (b). The blue curves represent weekly dengue cases, while the magenta dashed line denotes the long-term mean incidence used as the outbreak threshold (2010-2024). Open circles indicate threshold crossings that define outbreak intervals, black dots mark the outbreak peaks, and red vertical bars represent the detected events. The shaded region illustrates the tolerance window $\tau$ around an event. Peaks occurring within a $\pm 2$-week interval are considered temporally synchronized. }
	\label{fig1}
\end{figure}

As an illustrative example, Fig.~\ref{fig1}(a) and Fig.~\ref{fig1}(b) display weekly dengue incidence for two municipalities. In each panel, the magenta horizontal dotted line represents the long-term mean incidence ($\overline{c}_i$). An outbreak episode is defined as occurring when the time series crosses above this threshold; upward and downward crossings are indicated by open circles. This defines the outbreak interval $[t_{\rm start},\, t_{\rm end}]$. Within each interval, the week of maximum incidence (black dot) is identified and defined as the outbreak event. For instance, in panel (a), the first outbreak peak occurs in epidemiological week 14 of 2013, while in panel (b), the corresponding peak occurs in week 12 of 2013. Given a tolerance window of $\tau = \pm 2$ weeks, these two peaks are regarded as temporally synchronized. While this example illustrates synchronization between a pair of municipalities, the synchronization index defined in Eq.~(\ref{synchronization_eq}) generalizes this procedure to $N$ municipalities simultaneously, providing a state-level measure of outbreak alignment.

\section{Phase transition and geographical expansion of dengue}\label{phase_transition}
\begin{figure}[!ht]
	\centering
	\includegraphics[scale=0.37]{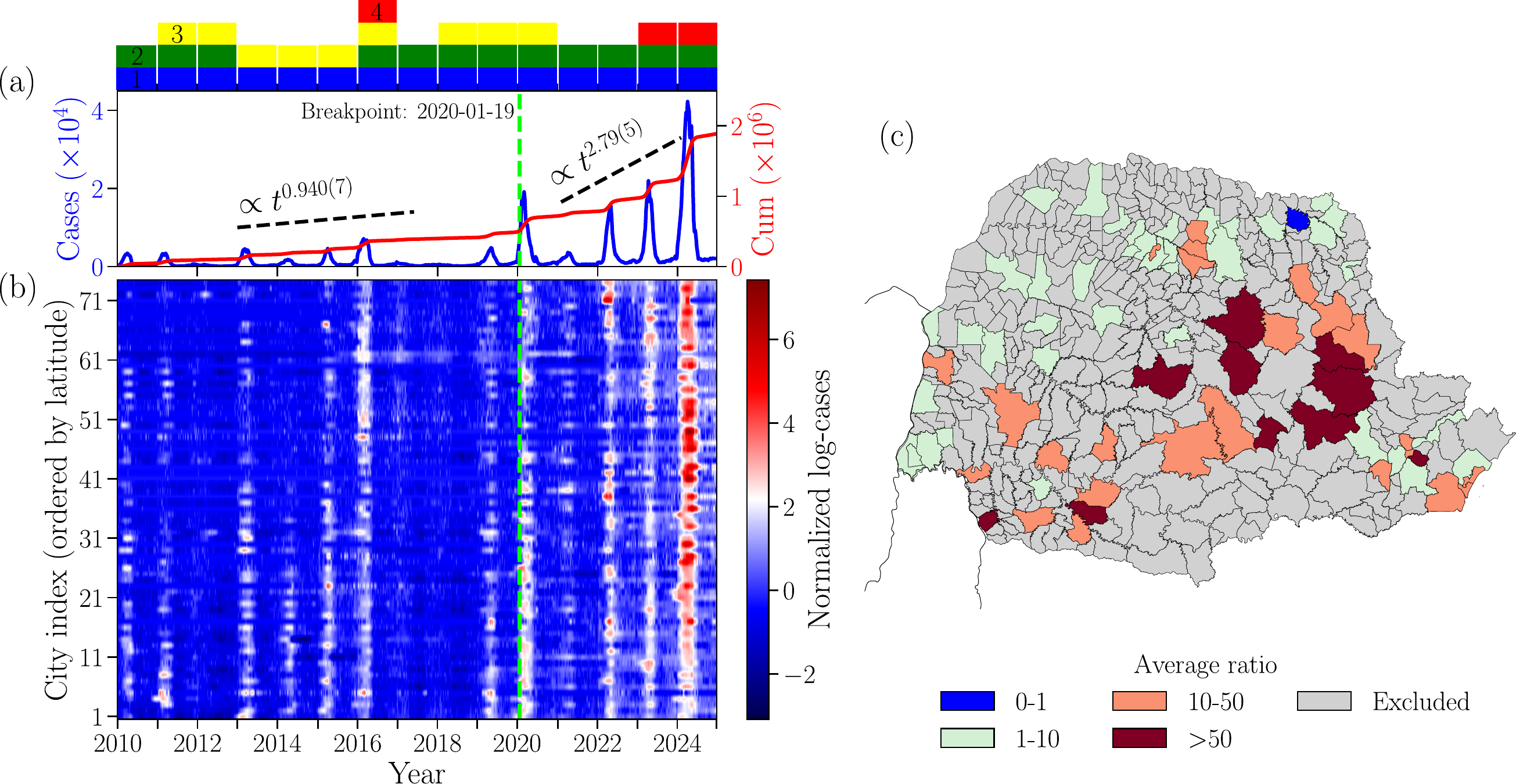}
	\caption{(a) Weekly notified dengue cases (blue curve) and cumulative cases (red curve). The serotype circulation in each year is displayed by the coloured blocks above panel (a), where the presence of DENV-1, 2, 3, and 4 is represented by blue, green, yellow, and red blocks. The lime vertical dashed line marks the estimated breakpoint (2020-01-19). Before and after the breakpoint, the cumulative curve follows different scaling regimes, growing as $t^{0.940(7)}$ and $t^{2.79(5)}$, respectively. (b) Heatmap of normalized dengue cases by year and city index (ordered by latitude). The normalization is defined as ${\rm Norm.\,Cases} = ({\rm cases}_{\log} - \overline{{\rm cases}_{\log}})/\sigma({\rm cases}_{\log})$, with ${\rm cases}_{\log} = \log({\rm cases}+1)$. (c) Map of Paran\'a showing the ratio between the average weekly number of dengue cases after and before the breakpoint; excluded municipalities are shown in gray.}
	
	\label{fig2}
\end{figure}

The notified dengue cases in Paran\'a display a clear temporal transition from a low-endemicity regime to a high-endemicity one (Fig. \ref{fig2}(a), blue curve). This transition is particularly evident in the cumulative time series (red curve in Fig. \ref{fig2}(a)), which reveals a sharp change in power-law behaviour, suggesting a changepoint at 2020-01-19 (vertical dotted lime line in Fig. \ref{fig2}(a)). Before this breakpoint, the cumulative incidence grows as $t^{0.940(7)}$, whereas after it accelerates sharply, proportional to $t^{2.79(5)}$. Several factors may contribute to this transition, including improvements in disease surveillance, which likely increased case detection, serotype circulation \cite{Aguiar2011}, human behaviour, and climate conditions. However, these factors alone do not readily explain the magnitude and spatial coherence of the observed increase, which extends beyond southern Brazil and is also observed across Latin America \cite{Almeida2025}. Furthermore, serotype co-circulation alone is not sufficient to explain the observed transition, because DENV-1, DENV-2, and DENV-3 co-circulated for most of the study period, with episodic reintroductions of DENV-4 in 2016, 2023, and 2024, as shown in the upper panel of Fig. \ref{fig2}(a). Notably, the largest outbreaks (2023 and 2024) coincide with DENV-4 reintroduction; nevertheless, serotype turnover alone does not account for the abrupt change in the scaling regime. Human behaviour is difficult to quantify, and relevant data are lacking; therefore, it is not investigated here. By contrast, climate conditions have been identified as an important factor in the expansion of dengue across different localities \cite{Xu2024,Cox2025}, and may help explain this abrupt transition. Together, these results suggest that the observed transition reflects a substantial change in transmission conditions, pointing to additional large-scale drivers, including climate variability and geographical expansion, as plausible contributors to the recent acceleration of dengue incidence.

The spatial patterns reveal a marked geographical expansion of dengue incidence, with cities previously unaffected, or reporting only a few cases, beginning to register cases after the breakpoint (Fig. \ref{fig2}(b)). Red and white tones indicate large positive deviations from the mean logarithmic number of cases, whereas blue tones indicate reductions. Before the breakpoint, most cities exhibit predominantly blue tones, with only localized white bands, mainly in northern municipalities. A first large-scale outbreak occurred in 2016, representing the most intense event up to that year, and affected practically all the considered cities. Following this event, a marked reduction in reported cases is observed in both Figs. \ref{fig2}(a) and \ref{fig2}(b), which may reflect partial population immunity. Despite this, a sharp resurgence of dengue occurred after 2020. In the post-breakpoint period, Fig. \ref{fig2}(b) displays widespread white and red regions, indicating large positive deviations from historical means. After 2022, several southern cities exhibited intense outbreaks, culminating in an unprecedented spatial expansion in 2024, which explains the extreme peak observed in that year. Relative to historical variability, 2024 represents an extreme event, with deviations exceeding seven standard deviations of the local time series. This spatial expansion substantially increases the burden on public health systems, as dengue transmission now affects cities that were previously unexposed.

By comparing the average weekly dengue cases before and after the breakpoint for each city, we observe a pronounced increase in incidence in almost all municipalities (Fig. \ref{fig2}(c)), with some cities exhibiting increases exceeding a factor of 50. These extreme ratios arise primarily because several cities were previously unaffected or reported only sporadic cases, resulting in near-zero average incidence before 2020. The most pronounced increases are concentrated in central-eastern municipalities. Beyond these critical areas, almost all cities display post-breakpoint ratios greater than unity, indicating a generalized rise in dengue transmission. Only one municipality exhibits a ratio between 0 and 1 (blue), which is associated with a highly noisy time series. Together, the patterns shown in Figs. \ref{fig2}(b) and \ref{fig2}(c) indicate that previous geographical limits of dengue transmission have been crossed, with all analysed cities now reporting cases. Notably, the year 2024 exhibits a continuous red band in Fig. \ref{fig2}(b), suggesting large-scale synchronization of outbreaks across the state.

The widespread spatial expansion and the emergence of nearly simultaneous outbreaks across municipalities suggest that dengue dynamics in Paran\'a are increasingly consistent with the influence of common external forcing, rather than isolated local processes.

\section{Spatiotemporal synchronization of dengue outbreaks}\label{synchronization}
The spatial diffusion of dengue incidence and the progressive reduction of local heterogeneities motivate the analysis of spatio-temporal synchronization as a defining feature of the post-breakpoint regime. As dengue expands geographically across the state, outbreaks that were previously asynchronous increasingly align in time, indicating a transition toward a more coherent dynamical regime.

\begin{figure}[!ht]
	\centering
	\includegraphics[scale=0.4]{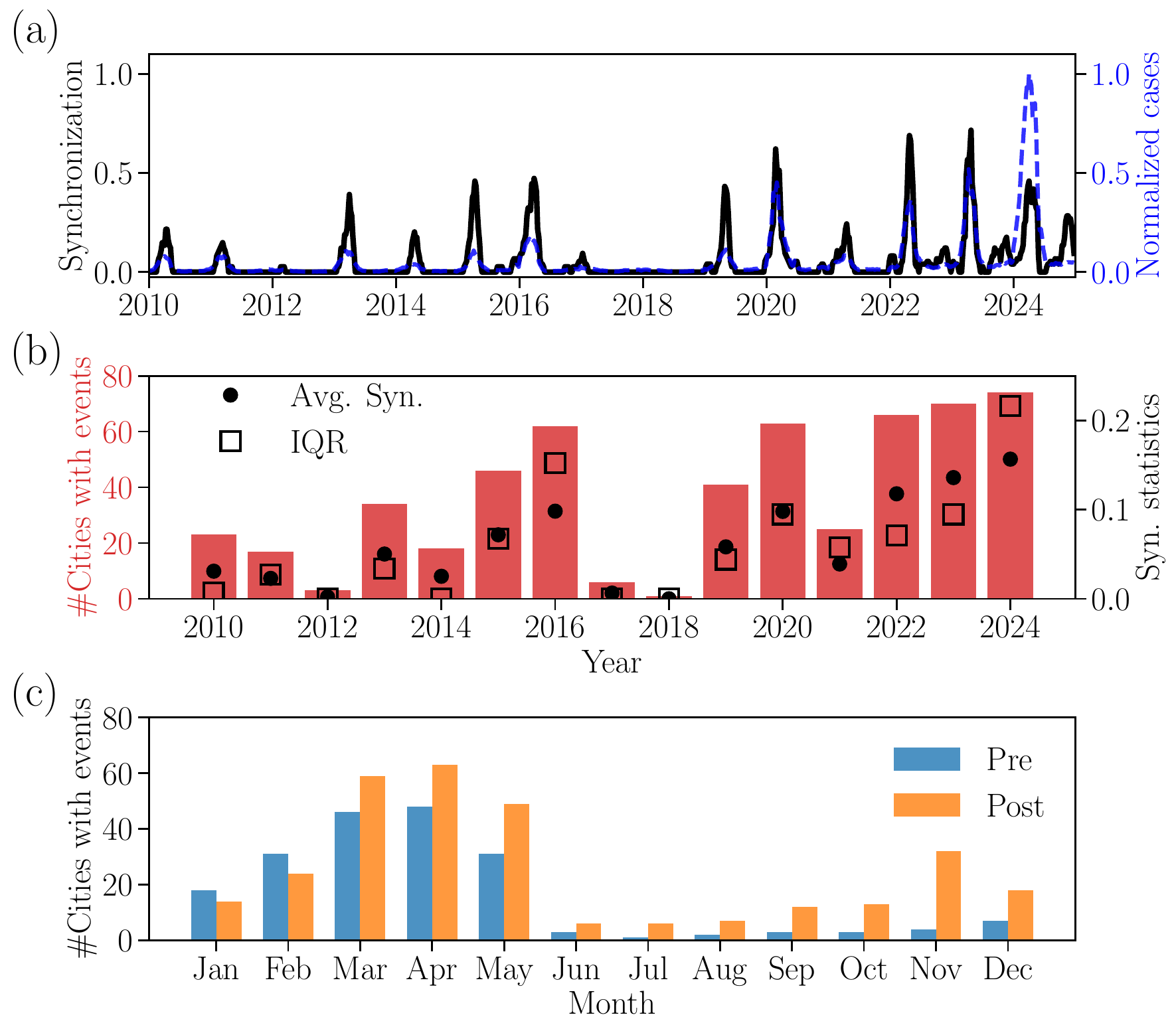}
	\caption{(a) Weekly synchronization (black curve) and fraction of dengue cases (blue dashed curve). (b) Red bars indicate the number of cities reporting at least one outbreak event per year, while black circles and open squares show the annual mean synchronization and interquartile range (IQR), respectively. (c) Number of cities reporting at least one outbreak event per month before and after the breakpoint, shown by the blue and orange bars, respectively. }
	\label{fig3}
\end{figure}

Employing the event synchronization metric, we find that the synchronization signal closely follows the fraction of dengue cases, as shown by the black and dotted blue curves, respectively, in Fig. \ref{fig3}(a). Synchronization peaks systematically coincide with peaks in incidence, resulting in a high correlation between the two signals (Spearman $\rho = 0.80$; Pearson $r = 0.68$). Overall, major outbreaks are characterized by elevated synchronization levels, with the notable exception of the 2024 outbreak. High peaks in the synchronization and fraction of cases indicate that large epidemics emerge from the quasi-simultaneous activation of multiple cities.

This interpretation is supported by Fig. \ref{fig3}(b), where the red bars represent the number of cities reporting at least one outbreak event per year. Years with high synchronization levels correspond to periods with a larger number of active cities (e.g., 2016, 2020, 2022, and 2023), whereas synchronization collapses when only a few municipalities report cases. Notably, 2012, 2017, and 2018 mark minima in the number of active cities and, consequently, in synchronization levels. The latter two years coincide with the period following the vaccination campaign initiated in 2016, which is not explored here.  Therefore, the number of cities reporting outbreaks provides essential context for interpreting the synchronization signal, as shown in Fig. \ref{fig3}(a).

Together, the information from Figs. \ref{fig3}(a) and \ref{fig3}(b) reveals four distinct synchronization regimes. \textbf{(i)} Between 2010 and 2016, dengue outbreaks exhibited strong and temporally concentrated synchronization, characterized by high synchronization peaks and low interquartile ranges (IQR; black squares in Fig. \ref{fig3}(b)), consistent with a well-defined seasonal pattern. During this period, the annual mean synchronization (black circles) also displays an increasing trend. \textbf{(ii)} This regime collapses in 2017--2018, when synchronization sharply decreases alongside a reduction in the number of active cities, reflecting the combined effects of epidemiological control and accumulated immunity. The number of active cities began to increase again after 2019. Despite the relatively small number of active cities, synchronization remains high. Overall, the pre-breakpoint regimes (\textbf{i} and \textbf{ii}) are characterized by limited spatial activation and synchronization constrained to a relatively fixed subset of cities.

\textbf{(iii)} After 2020, synchronization increases again, driven by a substantial rise in the number of active cities, except for 2021, which was strongly affected by the COVID-19 pandemic and disruptions in public health services. Up to 2023, the number of active cities, the fraction of cases, and the synchronization level remain strongly correlated, as evidenced by the signals in Figs. \ref{fig3}(a) and \ref{fig3}(b). \textbf{(iv)} In contrast, the 2024 outbreak exhibits a qualitatively different behaviour: although incidence reaches historically high levels, the maximum synchronization peak is lower than in previous years (e.g., 2023). At the same time, 2024 presents the highest annual mean synchronization and variability (IQR $= 0.21$). Unlike earlier outbreaks, synchronization in 2024 is no longer dominated by a single, temporally coherent event. Instead, multiple outbreak waves occur throughout the year across all cities, sustaining synchronization over an extended period. This pattern indicates a distinct synchronization regime, characterized by stronger temporal heterogeneity and overlapping epidemic waves.

This shift is further reflected in the seasonal distribution of outbreaks. Figure \ref{fig3}(c) shows the number of cities reporting at least one outbreak event per month before and after the breakpoint (blue and orange bars, respectively). In the pre-breakpoint regime, dengue activity is largely confined to the first five months of the year, whereas the post-breakpoint period exhibits a broader and temporally shifted activation pattern. Before the breakpoint, outbreaks peak in April, consistent with the historical seasonal cycle of temperature and precipitation. In contrast, after the breakpoint, outbreaks emerge earlier and persist later into the year, involving a substantially larger number of cities. This seasonal shift suggests a change in the climatic forcing acting on dengue transmission. Remarkably, sporadic outbreaks are now observed during winter months, when historical mean temperatures are below $20\,^\circ$C.

Overall, our results indicate that dengue synchronization in Paran\'a has transitioned from a temporally and spatially constrained regime, characterized by well-defined seasons and a limited set of endemic cities, to a more dispersed and persistent state, in which transmission involves newly exposed cities and extended activity periods. This transition is consistent with changes in climatic forcing and seasonality acting coherently across space.
\newpage

\section{Climate anomalies, permissive transmission days, and synchronization}\label{anomalia_sec}
As described above, dengue transmission in Paran\'a has undergone substantial changes in recent years, including a marked geographical expansion into previously unaffected or low-endemicity cities. Dengue dynamics are strongly constrained by environmental conditions that modulate the mosquito life cycle. Consequently, the number of days with climatically permissive conditions acts as an effective control parameter for transmission intensity. In this section, we show that anomalous weather conditions have substantially increased the number of conducive days, amplifying epidemic risk across the state.

\begin{figure}[!ht]
	\centering
	\includegraphics[scale=0.35]{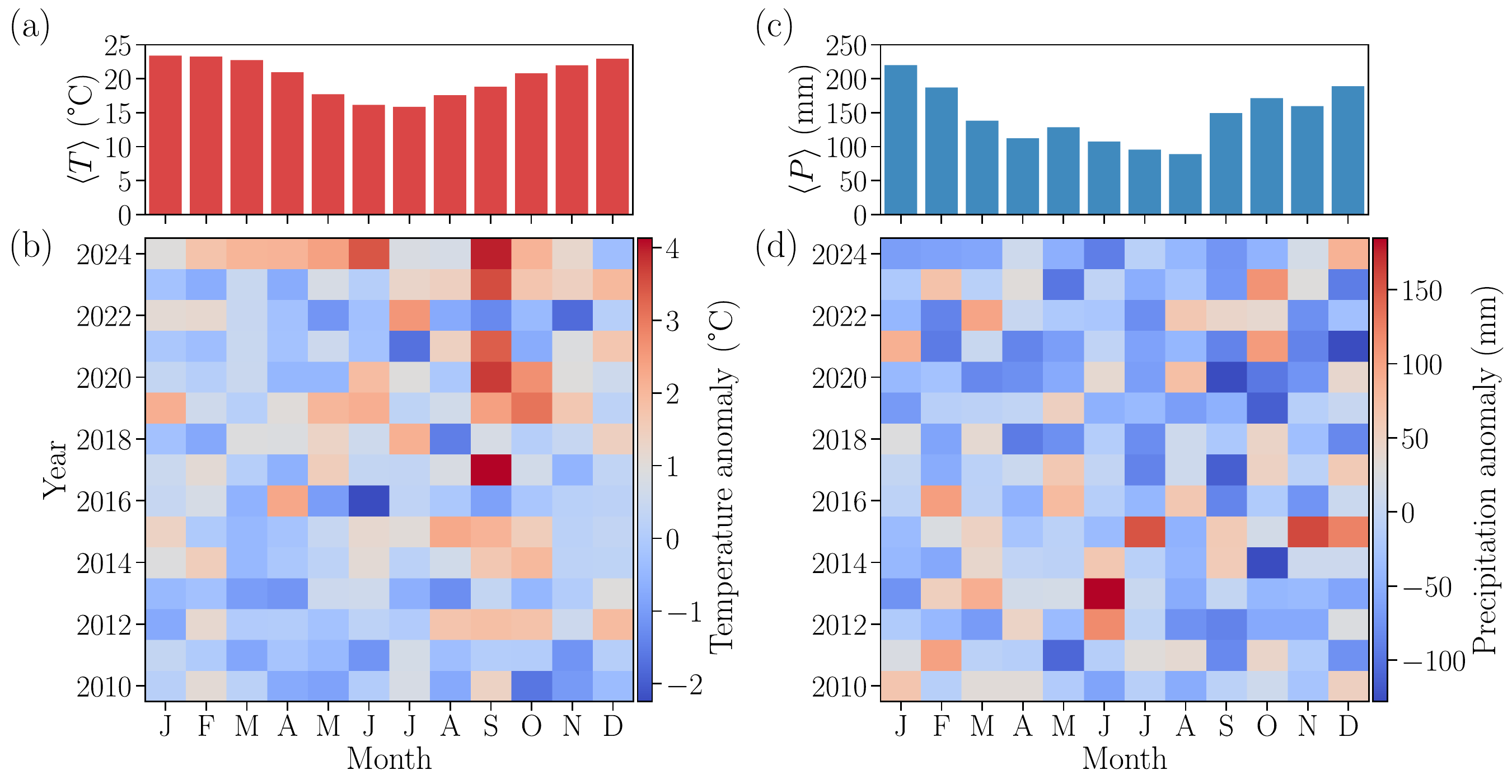}
	\caption{Panels (a) and (c) show the historical monthly mean temperature and precipitation in Paran\'a, respectively, based on the reference period 1980-01-01 to 2010-01-02. The notation $\langle \cdot \rangle$ means the spatial average across all considered municipalities. 
    Panels (b) and (d) show the corresponding monthly temperature and precipitation anomalies during the study period.}
	\label{fig4}
\end{figure}

Paran\'a exhibits four well-defined seasons, with monthly mean temperatures ranging from $15.85\,^\circ$C in July to $23.39\,^\circ$C in January (Fig. \ref{fig4}(a)), and a rainy season during the summer (Fig. \ref{fig4}(c)). This seasonal regime has historically confined dengue transmission to specific regions and periods of the year.

This historical climate pattern has been progressively modified by climate anomalies, as shown in Figs. \ref{fig4}(b) and \ref{fig4}(d) for temperature and precipitation, respectively. Temperature anomalies exhibit an increasing trend, particularly after 2017, with September showing the largest positive deviations. Furthermore, the years 2023 and 2024 are the warmest across almost all months, coinciding with the period of most intense dengue activity.

These thermal anomalies, together with the increase in temperatures observed since 2020, are consistent with the observed shift in the timing of outbreak events (Fig. \ref{fig3}(c)). Remarkably, 2024 is the year with record temperatures and dengue incidence, supporting the role of temperature as a key factor in dengue transmission.

Comparing these climatic characteristics with the four synchronization regimes identified in Sec. \ref{synchronization}, we find that regimes \textbf{(iii)} and \textbf{(iv)}, the most active ones, coincide with high temperature anomalies (Fig. \ref{fig4}(b)) and a redistribution of rainfall patterns (Fig. \ref{fig4}(d)). By contrast, during the collapse of cases and synchronization in regime \textbf{(ii)}, we do not observe relevant climate anomalies.

Precipitation patterns also show substantial deviations from their historical monthly means. While some months exhibit reduced rainfall (negative anomalies), others experience extreme rainfall events (Fig. \ref{fig4}(d)), indicating a reorganization of hydrological regimes across the state. In contrast to temperature, precipitation anomalies do not display a sustained increasing trend. Together, these results suggest a shift toward hotter conditions across most months, accompanied by more variable rainfall regimes, including both drought and extreme precipitation events.

\subsection{Anomalous temperature increases the number of thermal optimal days}
We define thermally possible days for transmission as those with temperatures between $18$ and $34\,^\circ$C, and thermally optimal days as those with temperatures between $26$ and $29\,^\circ$C \cite{Abdullah2022,Mordecai2017}. Based on these thresholds, we computed the monthly number of thermally permissive days for each city. Here, thermally permissive days refer to the combined set of possible and optimal thermal days. We then estimated the Pearson correlation between local temperature anomalies and the number of conducive thermal days. Averaged across the 74 cities, temperature anomalies exhibit a positive correlation with both optimal days ($r = 0.39$, $p = 0.01$) and possible days ($r = 0.26$, $p = 0.003$). These results indicate that temperature anomalies are associated with an increase in the number of thermally favourable days. Although the relationship includes non-linear components, a substantial linear component is still evident; therefore, we used linear regression to measure the effect of temperature anomaly on permissive days.

\begin{figure}[!ht]
	\centering
	\includegraphics[scale=0.33]{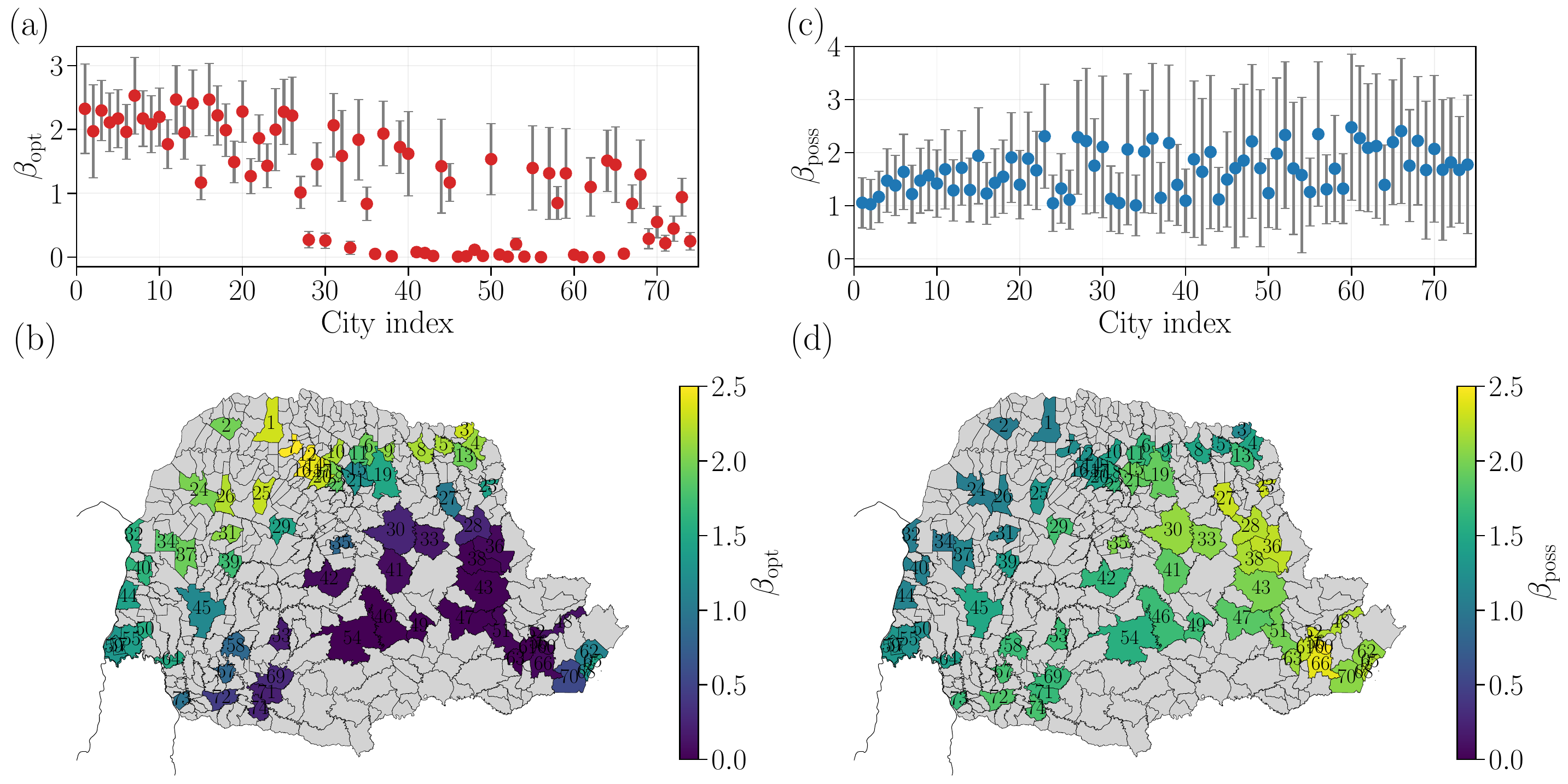}
	\caption{Regression coefficients from Eq.~\eqref{reg_linear_an} for (a,b) optimal and (c,d) possible thermal days. The regression quantifies the effect of temperature anomaly on the monthly number of thermally permissive days. Panels (a) and (c) show the coefficients as a function of city index, with error bars representing 95\% confidence intervals. Panels (b) and (d) show the corresponding spatial distributions.}
	\label{fig5}
\end{figure}

For each municipality, we fitted a linear regression of the form
\begin{equation}
y^i_t = \beta^i_0 + \beta^i_1 \, \mathrm{anomaly}^i_t + \epsilon^i_t, \label{reg_linear_an}
\end{equation}
where $i \in [1,74]$ denotes the municipality index, $y^i_t$ is the number of optimal or possible days per month, $\beta^i_{0,1}$ are regression coefficients, and $\epsilon^i_t$ represents random variability. When $\beta^i_1 > 0$ and $p < 0.05$, we interpret this as evidence of a positive association between temperature anomaly and thermally permissive days. Figures \ref{fig5}(a,b) and \ref{fig5}(c,d) display the regression coefficients for optimal and possible days, respectively. In both panels, the error bars indicate 95\% confidence intervals for $\beta^i_1$. For simplicity, we denote these coefficients by $\beta_{\rm opt}$ and $\beta_{\rm poss}$.

For northern cities (index 1 to 39), the number of thermally optimal days systematically increased with temperature anomaly (Figs. \ref{fig5}(a) and \ref{fig5}(b)). By contrast, southern municipalities display more heterogeneous responses, with many coefficients close to or equal to zero. Notably, several cities with $\beta_{\rm opt} \sim 0$ belong to a region that was strongly affected by dengue after 2020, as shown in Fig. \ref{fig3}(c). This apparent divergence suggests that optimal thermal conditions alone are not sufficient to explain recent outbreaks.

When the regression is performed with $y^i_t$ defined as the number of possible days, the results are shown in Figs. \ref{fig5}(c) and \ref{fig5}(d) reveal a consistent and statistically significant increase in $\beta_{\rm poss}$ across all considered cities. Remarkably, $\beta_{\rm poss}$ increases approximately linearly with city index, suggesting that temperature anomalies are expanding the number of possible days, especially in southern cities, which are historically cooler regions. Furthermore, Fig. \ref{fig5}(d) shows a cluster of municipalities with $\beta_{\rm poss} > 1.2$ that coincides with the municipalities most affected in Fig. \ref{fig3}(c). Thus, in cooler cities, anomalous warming does not always expand the number of days within the optimal transmission range, but it does increase the number of thermally possible days.

The regression coefficients are statistically significant, with a mean $p = 0.01$ for optimal days (63 cities with $p < 0.05$) and $0.003$ for possible days (all 74 cities with $p < 0.05$). These results reinforce that temperature anomalies are associated with increases in both optimal and possible thermal days.
Previous results established a robust relationship between temperature anomalies and thermally conducive days. We now assess whether these thermal conditions are associated with spatio-temporal synchronization.

As is well known, climate effects translate into dengue transmission with a delay. To estimate this delay between thermally favourable days and synchronization, we performed a cross-correlation analysis. Mean monthly synchronization was most strongly associated with the number of optimal days at a lag of 3 months ($r=0.44$). For possible days, the maximum correlation was achieved with a 1-month delay ($r=0.34$). Both delays are consistent with the literature \cite{Abdullah2022}. 

To quantify the contribution of thermal suitability to synchronization, we evaluated three regression models. First, we considered a linear model, given by
\begin{equation}
{\rm Syn}(t) = \beta_0 + \beta_1 y(t-\tau) + \epsilon(t), \label{linear_regression}
\end{equation}
where $y(t-\tau)$ denotes the climate predictor lagged by $\tau$. In this case, $y$ represents thermal permissibility. This model captures linear associations between synchronization and thermal indicators.

Since ${\rm Syn}(t)$ contains a large proportion of zero values ($\sim 46\%$), we used a hurdle model consisting of a logistic component for ${\rm Syn}(t)=0$, defined by
\begin{equation}
P({\rm Syn}(t) = 0 \mid y(t-\tau)) = \frac{1}{1 + e^{-(\beta_0 + \beta_1 y(t-\tau))}}, \label{logit_regression}
\end{equation}
and a Gamma generalized linear model with log link for ${\rm Syn}(t)>0$, given by 
\begin{equation}
\mathbb{E}\!\left[{\rm Syn}(t)\mid {\rm Syn}(t)>0, y(t-\tau)\right]
= \exp\!\big(\beta_0^{(G)} + \beta_1^{(G)} y(t-\tau)\big).  \label{gama_regression}
\end{equation}

For the three models, the regression results are summarized in Table \ref{tabela1}. The linear model reveals positive and statistically significant values of $\beta_1$, suggesting a positive association between thermal indicators and synchronization. Comparing both indicators, the number of optimal days has a stronger association with synchronization than the number of possible days. When both variables are included in a multiple regression, the linear model achieves $R^2 = 0.25$, higher than the simple-regression models. This suggests that combining both variables improves model fit, while preserving the stronger contribution of optimal days.

We then examined the nonlinear components of the hurdle model. The logistic component shows consistently negative values of $\beta_1$ with statistical significance for all three model specifications: optimal days, possible days, and the multiple-regression model. Again, optimal days show a stronger association than possible thermal days. These results indicate that increases in the number of thermally favourable days per month are associated with a reduced probability of observing asynchronous states. Therefore, conducive thermal conditions, with appropriate delays, are associated with the emergence of synchronization, with a dominant contribution from optimal days.

For observations with ${\rm Syn}(t)>0$, the Gamma model was employed. The results presented in Table \ref{tabela1} show positive and statistically significant parameters, suggesting a nonlinear association between thermal permissibility and the magnitude of synchronization. These results indicate that, once synchronization is present, thermally favourable days, with appropriate delays, are associated with higher synchronization levels. Again, optimal days show a stronger association than possible days.

Taken together, our findings support a two-stage mechanism linking temperature anomalies to dengue synchronization: $1)$ anomalous warming increases the number of thermally permissive days and is associated with a reduced probability of asynchronous states; and $2)$ once synchronization is present, appropriate thermal conditions are associated with higher synchronization levels. These results suggest that temperature may play a dual role in dengue synchronization, first favouring its emergence and subsequently sustaining the synchronous regime.

\begin{table}[htbp]
\centering
\caption{Estimated regression coefficients ($\beta_1$) for the association between thermal suitability indicators and synchronization. Results are shown for the linear model, the logistic component of the hurdle model for asynchronous states, and the Gamma component for positive synchronization values. Predictors correspond to the number of optimal (Opt.) and possible (Poss.) thermal days.}
\label{tabela1}
\begin{tabular}{l|c|c|c}
Model  & $y$ & $\beta_1$ [95\% CI] & $p$ \\
\hline
Linear & Opt. & 0.019 [0.013, 0.025] & $6.8\times10^{-10}$ \\ \hline
       & Poss. & 0.004 [0.002, 0.006] & $1.7\times10^{-6}$ \\ \hline
       & Opt. + Poss. &
       \begin{tabular}{c}
           0.017 [0.011, 0.023]\\
           0.003 [0.001, 0.005]
       \end{tabular}
       &
       \begin{tabular}{c}
           $8.7\times10^{-8}$\\
           $2.5\times10^{-4}$
       \end{tabular}
       \\ \hline \hline
Logit & Opt. & -0.396 [-0.622, -0.170] & $5.7\times10^{-4}$ \\ \hline
      & Poss. & -0.100 [-0.140, -0.056] & $5.5\times10^{-6}$ \\ \hline
      & Opt. + Poss. &
       \begin{tabular}{c}
           -0.311 [-0.529, -0.094]\\
           -0.080 [-0.123, -0.036]
       \end{tabular}
       &
       \begin{tabular}{c}
           $5.0\times10^{-3}$\\
           $3.0\times10^{-4}$
       \end{tabular}
       \\ \hline \hline
Gamma & Opt. & 0.106 [0.032, 0.180] & $4.9\times10^{-3}$ \\ \hline
      & Poss. & 0.082 [0.052, 0.112] & $6.0\times10^{-8}$ \\ \hline
      & Opt. + Poss. &
       \begin{tabular}{c}
           0.084 [0.019, 0.150]\\
           0.072 [0.042, 0.103]
       \end{tabular}
       &
       \begin{tabular}{c}
           $1.1\times10^{-2}$\\
           $3.5\times10^{-6}$
       \end{tabular}
       \\ \hline
\end{tabular}
\end{table}

\subsection{Anomalous precipitation and synchronization}
Precipitation is also an important factor in dengue transmission. To assess its effects, we classified rainy days according to the quantile technique \cite{Borges2025}, as shown in Table \ref{tabela2}. By computing the correlation between the number of days within each classification and precipitation anomaly, we found a high correlation between precipitation anomaly and very rainy days. A weak positive correlation was observed for rainy days, while weak to moderate negative correlations were found for the other categories. These results indicate that precipitation anomalies do not lead to a uniform increase in rainfall, but are instead associated with a redistribution of rainfall patterns, mainly characterized by more frequent very rainy days.

\begin{table}[h!]
\centering
\caption{Classification of precipitation regimes according to quantiles and correlation between the number of days within each classification and precipitation anomalies.}
\label{tabela2}
\vspace{0.2cm}
\begin{tabular}{l|c|c}
\hline
{Classification} & {Quantile (Q)} & Correlation (An.)\\
\hline
Very dry & $y \leq 0.15Q$ & -0.285\\
Dry  & $0.15Q < y \leq 0.35Q$ & -0.370\\
Optimal/Normal & $0.35Q < y \leq 0.65Q$ & -0.180\\
Rainy & $0.65Q \leq y < 0.85Q$ & 0.251\\
Very rainy & $y \geq 0.85Q$ & 0.755\\
\end{tabular}
\end{table}

\begin{figure}[!ht]
	\centering
	\includegraphics[scale=0.32]{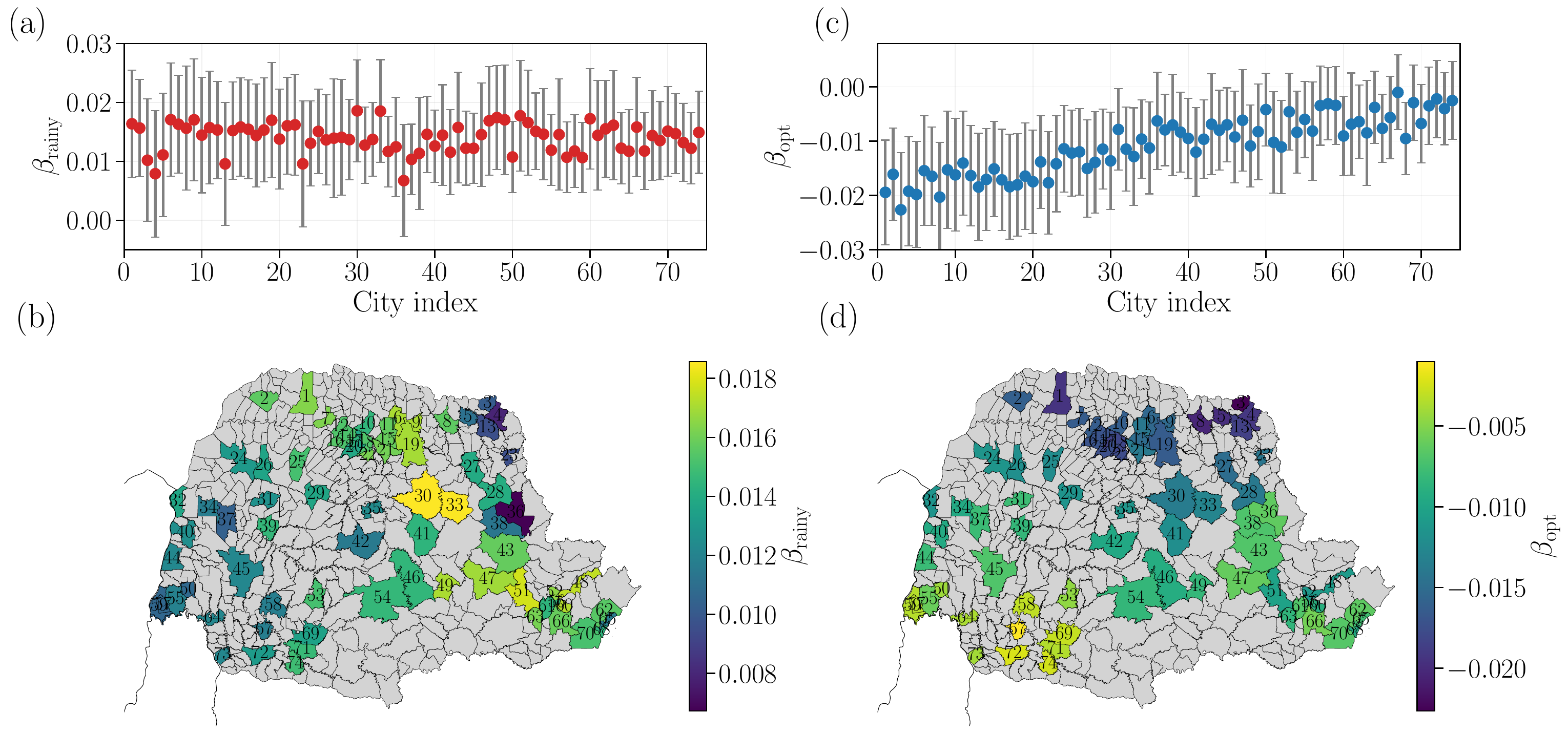}
	\caption{Regression coefficients from Eq.~\eqref{reg_linear_an} for (a,b) rainy and (c,d) optimal precipitation days. The regression quantifies the effect of precipitation anomaly on the monthly number of precipitation-defined days. Panels (a) and (c) show the coefficients as a function of city index, with error bars representing 95\% confidence intervals. Panels (b) and (d) show the corresponding spatial distributions.
    }
	\label{fig6}
\end{figure}

The observed redistribution of rainfall patterns is further supported by linear regression analyses (Eq. \eqref{linear_regression}), which yield positive coefficients for rainy days per month in all cities (Fig.~\ref{fig6}(a,b)). Although the magnitude of $\beta_{\rm rainy}$ remains relatively small ($\sim 0.01$), a small increase in the number of rainy days as a function of precipitation anomaly is evident. In contrast, the linear regression performed for optimal precipitation days shows negative coefficients across the cities (Fig.~\ref{fig6}(c,d)). Moreover, $\beta_{\rm opt}$ exhibits a spatial organization, increasing approximately linearly as a function of city index, suggesting that northern municipalities have fewer optimal precipitation days under positive anomalies. In southern municipalities, the number of optimal precipitation days remains practically unchanged.

The effects of precipitation on dengue synchronization are also expected to occur with a delay. Cross-correlation analysis shows that the strongest association between optimal precipitation days and synchronization occurs at a lag of 4 months ($r=0.210$), whereas rainy days show a stronger association at a lag of 2 months ($r=0.395$). These results suggest that optimal precipitation days are associated with synchronization at a longer lag and with a weaker correlation than rainy days.

Similarly to the temperature analysis, we performed regressions between precipitation variables and synchronization, with the results summarized in Table \ref{tabela3}. For the linear models, both variables yield positive and statistically significant coefficients ($p < 0.05$). The strongest association is observed for rainy days, which present a larger coefficient and a lower $p$-value than optimal precipitation days. This difference becomes more evident when both variables are included simultaneously in a multiple regression model.

We then performed the hurdle model analysis. In the logit component, the estimated coefficients for rainy days are negative and statistically significant, indicating that wet conditions are associated with a reduced probability of asynchronous states. In this analysis, rainy days emerge as the dominant variable, with a coefficient approximately eight times greater in magnitude than the coefficient associated with optimal precipitation days. This difference is also observed in the multiple regression model.

For observations with non-null synchronization, we used the Gamma model. When optimal and rainy variables are considered separately, both show positive coefficients, with a larger coefficient associated with optimal precipitation days. In the multiple regression model, however, rainy days become dominant, although the difference between coefficients is substantially reduced. This finding suggests that while rainy days are more strongly associated with the emergence of synchronization, both precipitation variables contribute to maintaining higher synchronization levels.

\begin{table}[htbp]
\centering
\caption{
Estimated regression coefficients ($\beta_1$) obtained from the linear model (Eq. \eqref{linear_regression}) and from the two components of the hurdle model: the logistic regression for $P({\rm Syn}=0)$ (Eq. \eqref{logit_regression}) and the Gamma generalized linear model with log link for $\mathbb{E}[{\rm Syn}\mid {\rm Syn}>0]$ (Eq. \eqref{gama_regression}). Predictors correspond to the number of optimal (Opt.) and rainy precipitation days per month, while the response variable is the synchronization index.}
\label{tabela3}
\vspace{0.2cm}
\begin{tabular}{l|c|c|c}
Model  & $y$ & $\beta_1$ [95\% CI] & $p$ \\
\hline
Linear & Opt. & 0.006 [0.002, 0.010] & $5.0\times10^{-3}$ \\ \hline
       & Rainy & 0.011 [0.007, 0.015] & $4.4\times10^{-8}$ \\ \hline
       & Opt. + Rainy &
       \begin{tabular}{c}
           0.005 [0.001, 0.008]\\
           0.011 [0.007, 0.015]
       \end{tabular}
       &
       \begin{tabular}{c}
           $2.0\times10^{-2}$\\
           $3.0\times10^{-7}$
       \end{tabular}
       \\ \hline \hline
Logit & Opt. & -0.027 [-0.103, 0.048] & $4.7\times10^{-1}$ \\ \hline
      & Rainy & -0.220 [-0.318, -0.121] & $1.1\times10^{-5}$ \\ \hline
      & Opt. + Rainy &
       \begin{tabular}{c}
           -0.005 [-0.085, 0.074]\\
           -0.214 [-0.314, -0.115]
       \end{tabular}
       &
       \begin{tabular}{c}
           $8.9\times10^{-1}$\\
           $2.4\times10^{-5}$
       \end{tabular}
      \\ \hline \hline
Gamma & Opt. & 0.091 [0.034, 0.149] & $1.6\times10^{-3}$ \\ \hline
      & Rainy & 0.075 [0.021, 0.130] & $6.6\times10^{-3}$ \\ \hline
      & Opt. + Rainy &
       \begin{tabular}{c}
           0.080 [0.024, 0.133]\\
           0.118 [0.061, 0.174]
       \end{tabular}
       &
       \begin{tabular}{c}
           $4.4\times10^{-3}$\\
           $3.8\times10^{-5}$
       \end{tabular}
       \\ \hline
\end{tabular}
\end{table}
\subsection{Multiple regression model -- temperature and precipitation}
The previous results suggest that temperature and precipitation are associated with a two-stage mechanism in synchronization, whereby they are linked to a reduced likelihood of asynchronous states and, subsequently, to the maintenance of coherent activity across the network. We now consider regression models including all four climatic variables, exemplified by
\begin{eqnarray}
{\rm Syn} (t) &=& \beta_0 + \beta_1 \, {\rm opt.\ temp.}(t-{\rm lag}_1) + \beta_2 \, {\rm poss.\ temp.}(t-{\rm lag}_2) \nonumber \\ 
&+& \beta_3 \, {\rm opt.\ prec.}(t-{\rm lag}_3) + \beta_4 \, {\rm rainy}(t-{\rm lag}_4) + \epsilon(t), \label{final_model}
\end{eqnarray}
where $\beta_1$ and $\beta_2$ are associated with optimal and possible temperature, respectively, while $\beta_3$ and $\beta_4$ correspond to optimal precipitation and rainy days. All variables are considered with their respective delays.

To remain consistent with the previous analyses, we estimate the associated parameters using linear, logistic, and Gamma models, with the results presented in Table \ref{tabela4}. Overall, the multiple regression framework provides a more comprehensive description of the observed synchronization patterns.

All models capture distinct contributions of each variable, with the strongest associations observed for the number of rainy days and optimal temperature. The former is related to the availability of breeding sites for mosquito populations, while the latter reflects favourable thermal conditions for mosquito development and virus transmission.

The logistic component shows negative and statistically significant coefficients for most variables, indicating that increases in the number of climatically favourable days are associated with a reduced probability of asynchronous states. This result is consistent with previous findings.

For observations with ${\rm Syn}(t)>0$, the Gamma model shows positive and statistically significant coefficients, suggesting that climatic conditions are associated with higher levels of synchronization once it is present.

Taken together, these results are consistent with a two-stage mechanism in which climatic conditions are associated with both the emergence and persistence of synchronization. These findings highlight the role of climate variability in shaping the collective dynamics of dengue outbreaks. Additionally, it is important to note that such favourable climatic conditions may become more frequent under climate change, particularly in temperate regions.

\begin{table}[h!]
\centering
\caption{Multivariate climate--synchronization regression results. 
Estimates are reported as $\beta$ [95\% CI] with corresponding $p$ 
for the linear model, the logistic component (probability of an asynchronous state), 
and the Gamma component (magnitude of synchronization conditional on ${\rm Syn}>0$). 
The parameters $\beta_1$, $\beta_2$, $\beta_3$, and $\beta_4$ are associated with optimal temperature (lag of 3 months), possible temperature (lag of 1 month), optimal precipitation (lag of 4 months), and rainy days (lag of 2 months), respectively.}
\label{tabela4}
\vspace{0.2cm}
\resizebox{\textwidth}{!}{%
\begin{tabular}{c|c|c|c|c|c|c}
\hline
 & \multicolumn{2}{c|}{{Linear}} 
 & \multicolumn{2}{c|}{{Logit}} 
 & \multicolumn{2}{c}{{Gamma}} \\
\hline
 & $\beta$ [95\% CI] & $p$ 
 & $\beta$ [95\% CI] & $p$ 
 & $\beta$ [95\% CI] & $p$ \\
\hline
$\beta_1$ & 0.014 [0.008, 0.020] & $8.09\times10^{-6}$ 
          & -0.300 [-0.528, -0.071] & $1.0\times10^{-2}$ 
          & 0.064 [0.000, 0.128] & $4.8\times10^{-2}$ \\

$\beta_2$ & 0.003 [0.001, 0.005] & $5.4\times10^{-3}$
          & -0.057 [-0.105, -0.010] & $1.8\times10^{-2}$ 
          & 0.051 [0.018, 0.083] & $2.0\times10^{-3}$ \\

$\beta_3$ & 0.004 [0.001, 0.008] & $2.2\times10^{-2}$
          & -0.011 [-0.104, 0.082] & $8.1\times10^{-1}$ 
          & 0.057 [0.005, 0.109] & $3.0\times10^{-2}$ \\

$\beta_4$ & 0.006 [0.002, 0.011] & $2.6\times10^{-3}$ 
          & -0.134 [-0.241, -0.026] & $1.5\times10^{-2}$ 
          & 0.076 [0.018, 0.134] & $9.7\times10^{-3}$ \\
\hline
\end{tabular}
}
\end{table}
\section{Validation in different localities}\label{validation}
To explore the generality and limitations of the two-stage mechanism, we repeated the analysis for two additional Brazilian states: Cear\'a and Minas Gerais, located in the Northeast and Southeast regions, respectively. Both are endemic regions for dengue.

First, we estimated the climatic lags for the 81 selected municipalities out of 184 in Cear\'a using cross-correlation analysis. Optimal temperature exhibited a maximum correlation of 0.52 with synchronization at a lag of 6 months, whereas possible temperature reached a weaker maximum correlation of 0.19 at 8 months. For precipitation variables, optimal precipitation showed a moderate correlation of 0.265 at a lag of 4 months, while the number of rainy days displayed a substantially stronger correlation of 0.650 at a lag of 14 months. These lag structures suggest distinct temporal pathways linking climatic exposure to epidemic coherence. While thermal suitability appears to be associated with synchronization on intermediate time scales (6--8 months), rainfall, particularly rainy days, operates over longer temporal windows, potentially reflecting cumulative ecological effects on vector population dynamics.

We then incorporated these lags into the multivariate regression framework (Table \ref{tabela5}). In the linear model, optimal temperature ($\beta_1$), optimal precipitation ($\beta_3$), and rainy days ($\beta_4$) are positively and significantly associated with synchronization, with rainy days showing the strongest association ($p < 10^{-12}$). In contrast, possible temperature ($\beta_2$) does not exhibit a significant linear contribution.

The logistic component does not reveal statistically significant predictors, indicating that in Cear\'a, climate variables are not strongly associated with the probability of asynchronous states. This contrasts with more transitional settings, where climatic permissibility appears to be associated with the emergence of coherent epidemic activity, and suggests that baseline climatically conducive days in Cear\'a may already sustain persistent epidemic coherence.

The Gamma regression reveals a different pattern. Conditional on synchronization being present (${\rm Syn} > 0$), both optimal temperature and rainy days are significantly associated with higher synchronization magnitude, with highly robust effects ($p < 10^{-7}$ and $p < 10^{-13}$, respectively). This indicates that, although climate does not appear to be associated with the onset of synchronization in Cear\'a, it is strongly associated with its intensity.

Taken together, these results point to a regime characterized by persistent baseline synchronization, where climatic variability appears to act primarily as an amplifier rather than as a trigger. This behaviour is consistent with the semi-arid tropical climate of Cear\'a, where thermal suitability is maintained for most of the year and rainfall pulses may generate delayed and cumulative effects on vector abundance.

\begin{table}[h!]
\centering
\caption{Multivariate climate--synchronization regression results for Cear\'a. 
Estimates are reported as $\beta$ [95\% CI] with corresponding $p$ 
for the linear model, the logistic component (probability of an asynchronous state), 
and the Gamma component (magnitude of synchronization conditional on ${\rm Syn}>0$). The parameters $\beta_1$, $\beta_2$, $\beta_3$, and $\beta_4$ are associated with optimal temperature (lag of 6 months), possible temperature (lag of 8 months), optimal precipitation (lag of 4 months), and rainy days (lag of 14 months), respectively.}
\label{tabela5}
\vspace{0.2cm}
\resizebox{\textwidth}{!}{%
\begin{tabular}{c|c|c|c|c|c|c}
\hline
\multicolumn{7}{c}{Cear\'a} \\
\hline
 & \multicolumn{2}{c|}{Linear} 
 & \multicolumn{2}{c|}{Logit} 
 & \multicolumn{2}{c}{Gamma} \\
\hline
 & $\beta$ [95\% CI] & $p$ 
 & $\beta$ [95\% CI] & $p$ 
 & $\beta$ [95\% CI] & $p$ \\
\hline
$\beta_1$ & 0.003 [0.001, 0.004] & $4.0\times10^{-4}$ 
          & 0.014 [-0.078, 0.107] & $7.5\times10^{-1}$
          & 0.032 [0.020, 0.043] & $7.3\times10^{-8}$ \\

$\beta_2$ & 0.005 [-0.011, 0.020] & $5.4\times10^{-1}$
          & -0.616 [-1.695, 0.463] & $2.6\times10^{-1}$
          & 0.077 [-0.049, 0.203] & $2.3\times10^{-1}$ \\

$\beta_3$ & 0.003 [0.001, 0.005] & $7.0\times10^{-4}$
          & -0.163 [-0.394, 0.069] & $1.6\times10^{-1}$ 
          & 0.013 [-0.003, 0.029] & $1.1\times10^{-1}$ \\

$\beta_4$ & 0.010 [0.007, 0.012] & $1.12\times10^{-12}$
          & -0.688 [-1.605, 0.229] & $1.4\times10^{-1}$ 
          & 0.075 [0.055, 0.095] & $1.5\times10^{-13}$ \\
\hline
\end{tabular}
}
\end{table}

Compared with Cear\'a, the 180 selected municipalities from Minas Gerais exhibit a different structure in the climate--synchronization relationship (Table \ref{tabela6}). In the linear component, all predictors are positively and significantly associated with synchronization, indicating that both thermal and precipitation suitability are associated with coherent epidemic dynamics across cities. In particular, optimal temperature ($\beta_1$) and rainy days ($\beta_4$) display the strongest associations, suggesting that favourable thermal conditions and rainfall are linked to spatial epidemic coherence. For this model, we obtained $R^2=0.38$.

The main difference compared with Cear\'a appears in the logistic component. In Minas Gerais, both possible temperature ($\beta_2$) and rainy days ($\beta_4$) are statistically significant and negative. This indicates that climatic permissibility is associated with a reduced likelihood of asynchronous epidemic activity. In this sense, adequately lagged meteorological conditions are associated with increased coherence among outbreaks across the state. This result is similar to that observed in Paran\'a, where the components $\beta_1$, $\beta_2$, and $\beta_4$ are associated with a reduced probability of asynchronous states.

Once synchronization is present, the Gamma regression results in Table \ref{tabela6} indicate that most climatic variables are associated with higher synchronization levels, except $\beta_1$, whose confidence interval crosses zero. These results suggest that climate in Minas Gerais is associated not only with the emergence of synchronized states, but also with the maintenance of spatial coherence.

\begin{table}[h!]
\centering
\caption{Multivariate climate--synchronization regression results for Minas Gerais. 
Estimates are reported as $\beta$ [95\% CI] with corresponding $p$ 
for the linear model, the logistic component (probability of an asynchronous state), 
and the Gamma component (magnitude of synchronization conditional on ${\rm Syn}>0$). The parameters $\beta_1$, $\beta_2$, $\beta_3$, and $\beta_4$ are associated with optimal temperature (lag of 4 months), possible temperature (lag of 4 months), optimal precipitation (lag of 12 months), and rainy days (lag of 1 month), respectively.}
\label{tabela6}
\vspace{0.2cm}
\resizebox{\textwidth}{!}{%
\begin{tabular}{c|c|c|c|c|c|c}
\hline
\multicolumn{7}{c}{Minas Gerais} \\
\hline
 & \multicolumn{2}{c|}{Linear} 
 & \multicolumn{2}{c|}{Logit} 
 & \multicolumn{2}{c}{Gamma} \\
\hline
 & $\beta$ [95\% CI] & $p$ 
 & $\beta$ [95\% CI] & $p$ 
 & $\beta$ [95\% CI] & $p$ \\
\hline
$\beta_1$ & 0.014 [0.007, 0.021] & $8.0\times10^{-5}$ 
          & -0.813 [-2.320, 0.694] & $2.9\times10^{-1}$
          & 0.088 [-0.003, 0.179] & $5.7\times10^{-2}$ \\

$\beta_2$ & 0.005 [0.002, 0.008] & $1.8\times10^{-3}$
          & -0.215 [-0.328, -0.103] & $1.6\times10^{-4}$
          & 0.111 [0.037, 0.185] & $3.3\times10^{-3}$ \\

$\beta_3$ & 0.004 [0.001, 0.007] & $1.3\times10^{-2}$
          & -0.005 [-0.111, 0.101] & $9.2\times10^{-1}$ 
          & 0.058 [0.012, 0.103] & $1.3\times10^{-2}$ \\

$\beta_4$ & 0.008 [0.005, 0.011] & $4.9\times10^{-8}$
          & -0.279 [-0.396, -0.161] & $3.2\times10^{-6}$ 
          & 0.113 [0.069, 0.157] & $5.7\times10^{-7}$ \\
\hline
\end{tabular}
}
\end{table}
\section{Sensitivity analysis of the synchronization window and outbreak threshold}\label{sensitivity_analysis}
We performed a sensitivity analysis to quantify how the synchronization window ($\tau$) and outbreak threshold affect the average synchronization over the entire time series, denoted by $\overline{\rm Syn}$.

In the main analysis, outbreaks were defined using the condition $c_i(t) \geq \overline{c}_i + 1$. To explore a broader range of outbreak definitions, we generalized this threshold to $c_i(t) \geq \overline{c}_i + 1 + k\,\sigma_i$, where $\sigma_i$ is the standard deviation of each municipal time series and $k$ is a constant.

As a reference, we considered the average synchronization obtained for $\tau=2$ and $k=0$. The relative impact of each parameter combination was then defined as
\begin{equation}
\text{impact}(\tau,k) = \left(\overline{\rm Syn}(\tau,k) - \overline{\rm Syn}(2,0)\right)\times 100\%.
\end{equation}

\begin{figure}[!ht]
	\centering
	\includegraphics[scale=0.45]{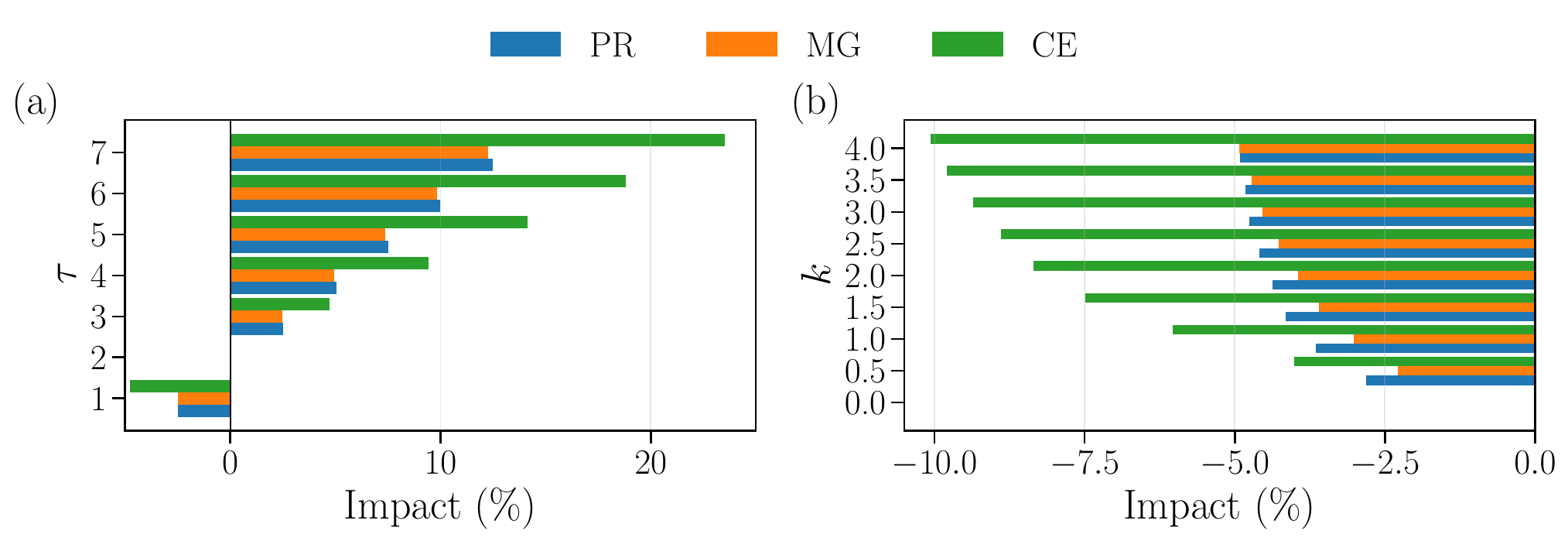}
	\caption{Sensitivity analysis of the average synchronization $\overline{\rm Syn}$ with respect to the synchronization window $\tau$ and the outbreak threshold parameter $k$. The impact is defined relative to the baseline case $(\tau=2, k=0)$ as $\text{impact}(\tau,k) = \left(\overline{\rm Syn}(\tau,k) - \overline{\rm Syn}(2,0)\right)\times 100\%$. (a) Impact of varying the synchronization window $\tau$ while fixing $k=0$. (b) Impact of varying the threshold parameter $k$ while fixing $\tau=2$. Results are shown for Paran\'a (PR), Minas Gerais (MG), and Cear\'a (CE) in blue, orange, and green, respectively.}
	\label{fig7}
\end{figure}

First, fixing $k=0$, we evaluated the effect of varying $\tau$. The results, shown in Fig.~\ref{fig7}(a), indicate that for $\tau=1$ the average synchronization decreases, whereas for $\tau>2$ it increases monotonically. For all three states, Paran\'a (PR), Minas Gerais (MG), and Cear\'a (CE), the impact grows approximately linearly with $\tau$. This behaviour is expected, as the synchronization measure is defined through a temporal window that accumulates coincident events, leading to an inherent dependence on $\tau$.

Next, fixing $\tau=2$, we analysed the effect of varying the threshold parameter $k$, as shown in Fig.~\ref{fig7}(b). In this case, the dependence is nonlinear and exhibits a systematic decrease in synchronization as $k$ increases. This reflects the progressive filtering of weaker outbreaks, which reduces the number of detected events and, consequently, the likelihood of temporal coincidence across cities.

Importantly, despite these quantitative changes, the qualitative patterns remain consistent across parameter values. In particular, Cear\'a exhibits stronger sensitivity to both $\tau$ and $k$, while Paran\'a and Minas Gerais display similar responses. These results indicate that our main findings are robust with respect to the choice of synchronization window and outbreak definition.

\section{Conclusions}\label{conclusions}
In this study, we systematically analysed the long-term transmission of dengue fever across 74 municipalities in Paran\'a, Brazil, between 2010 and 2024. Historically, this subtropical state exhibited relatively low dengue incidence compared with other Brazilian states, with cases spatially restricted to a limited subset of cities. Our analyses reveal a clear epidemiological transition after 2020, marking a shift from a low-endemicity regime to a high-endemicity state characterized by the expansion of dengue into previously less affected municipalities.

The post-breakpoint regime is characterized by a pronounced geographic expansion of dengue and by a substantial change in the spatiotemporal pattern of outbreaks. Dengue transmission becomes increasingly synchronized across cities, reflecting a transition from localized, seasonally constrained epidemics to a more coherent dynamical state. Although high synchronization typically coincides with large outbreaks, the unprecedented epidemic of 2024 exhibits a distinct pattern: synchronization is sustained over long periods through multiple overlapping waves, rather than being dominated by a single seasonal peak. This shift is accompanied by a substantial change in the dengue season, with outbreaks emerging earlier, persisting longer, and occurring even during winter months.

To investigate potential drivers of this transition, we examined the role of climatic forcing, focusing on temperature and precipitation anomalies. We show that anomalous weather conditions are associated with a substantial increase in the number of thermally and hydrologically conducive days for dengue transmission. Importantly, these increases are not uniform but reflect a reshaping of climatic conditions, with warming and intensified rainfall extremes.

Our results support a two-stage mechanism linking climate anomalies to dengue synchronization. First, the expansion of climate-conducive days is associated with a reduced probability of asynchronous states and with the emergence of coherent transmission across cities. Once synchronization is established, climatic suitability is further associated with higher and more persistent synchronization through delayed nonlinear effects. This mechanism provides a useful framework for connecting climate variability, geographical expansion, and large-scale epidemic coherence.

Taken together, our findings suggest that changing climatic conditions may be reshaping the dynamics of dengue in Paran\'a, contributing to the transformation of a spatially fragmented and seasonal disease into a more persistent and regionally synchronized public health threat. Beyond its regional relevance, this study highlights how gradual climatic changes may contribute to abrupt epidemiological transitions by altering the collective dynamics of disease spread.

To assess the robustness of our findings across different climatic regimes, we extended the analysis to two additional Brazilian states, Cear\'a and Minas Gerais. Comparing the results for the three states, we find a consistent pattern in the linear component, with statistically significant associations for most predictors across all regions. However, the proposed two-stage mechanism is not universally expressed. The logistic regression indicates that conducive climatic conditions are significantly associated with a reduced probability of asynchronous epidemic states in Paran\'a and Minas Gerais, whereas this effect is not observed in Cear\'a. We argue that this difference may be associated with the persistent climatic suitability in Cear\'a, where temperatures remain within the optimal or possible range for mosquito development during most of the year. Under such conditions, synchronization may not require an additional climatic trigger. In contrast, the Gamma component, which captures the magnitude and maintenance of synchronization once it is established, is statistically significant in all three states. This suggests that favourable climatic conditions consistently enhance the temporal persistence and spatial coherence of dengue outbreaks, even though their role in the emergence of synchronization may depend on the regional climatic context.

Our results highlight the important role of climate conditions in shaping dengue synchronization, acting through a two-stage mechanism: first, favouring the emergence of synchronous regimes and subsequently sustaining them. These findings underscore the need to incorporate climate-sensitive synchronization indicators into dengue surveillance, early-warning systems, and adaptation strategies, particularly in temperate and subtropical regions that are becoming increasingly vulnerable under ongoing climate change.

\enlargethispage{20pt}

\section*{Data}
All code and processed datasets used in this study are openly available on Zenodo \cite{GabrickZenodo} and GitHub \cite{GabrickGithub}.

\section*{Acknowledgements}
I.L.C is funded by CNPq (No.~302665/2017-0),  Coordena\c{c}\~ao de Aperfei\c{c}oamento 
de Pessoal de N\'ivel Superior (CAPES), and FAPESP (No.~2024/05700-5). 
E.C.G.~acknowledges financial support from FAPESP under grants No.~2025/02318-5, 
and  2025/24097-0. A.M.B. acknowledges financial support from 
CNPq, CAPES, and FAPESP (No.~2025/14690-6). M.A. acknowledges financial support from the Spanish Ministerio de Ciencia e Innovación (MICINN) through the Ramón y Cajal grant RYC2021-031380-I. This research was supported by the Basque Government through the “Mathematical Modeling Applied to Health” project (BERC 2022–2025 programme) and by the Spanish Ministry of Science, Innovation and Universities through the BCAM Severo Ochoa accreditation CEX2021-001142-S (MICINN/AEI/10.13039/501100011033).


\bibliographystyle{elsarticle-num}
\bibliography{Ref}

\end{document}